\documentclass[11pt,fleqn]{article}

\usepackage{amsfonts}
\usepackage{amsmath}
\usepackage{amssymb}
\usepackage{mathtools}
\usepackage{mathrsfs}
\usepackage{enumerate}
\usepackage{bbm}

\usepackage{pstricks}
\usepackage{subfig}
\usepackage{graphicx}
\usepackage{units}

\usepackage{caption}
\usepackage{booktabs}
\usepackage{textcomp}
\usepackage{array}
\usepackage{cite}
\usepackage{cancel}

\date{\today}

\newcommand{\del}{\partial}                            
\newcommand{\ol}[1]{\overline{#1}}                     
\newcommand{\abs}[1]{ \left| #1 \right| }              
\newcommand{\klammer}[1]{\left(#1\right)}              
\newcommand{\real}{\Re\mathfrak{e}\hspace{1pt}}        
\newcommand{\s}[1]{\texttt{#1}}                        

\def\ov{\overline}

\newcolumntype{z}[1]{>{\RaggedRight\hspace{0pt}}p{#1}}
\newcolumntype{w}[1]{>{\RaggedRight\hspace{0pt}}p{#1}}
\newcolumntype{v}[1]{>{\Centering\hspace{0pt}}p{#1}}

\DeclareCaptionLabelSeparator{mysep}{\hspace{3pt}:\hspace{3pt}}
\DeclareCaptionLabelFormat{mypiccap}{Fig.\hspace{3pt}{#2}}
\DeclareCaptionLabelFormat{mytabcap}{Tab.\hspace{3pt}{#2}}

\begin{document}

\date{}
\title{
\vskip 2cm
{\bf\huge Gaugino Condensation with\\
a Doubly Suppressed Gravitino Mass}\\[0.8cm]}

\author{{\sc\normalsize
Val\'eri L\"owen\footnote{E-mail: loewen@th.physik.uni-bonn.de}, Hans Peter Nilles\footnote{E-mail: nilles@th.physik.uni-bonn.de}, Andrea Zanzi\footnote{E-mail: zanzi@th.physik.uni-bonn.de}\!
\!}\\[1cm]
{\normalsize Physikalisches Institut der Universit\"at Bonn}\\
{\normalsize Nussallee 12, 53115 Bonn, Germany}\\[1cm]
}
\maketitle \thispagestyle{empty}
\begin{abstract}
{Supersymmetry breakdown via gaugino condensation in heterotic
string theory can lead to models with a doubly suppressed gravitino
mass. A TeV scale gravitino can emerge from a condensate as large
as the grand unified scale. We analyze the properties of these models
and discuss applications for particle physics and cosmology.
}
\end{abstract}

\clearpage

\tableofcontents
\newpage

\section{Introduction}

One of the basic questions in the study of phenomenological properties
of string theories concerns the origin of supersymmetry breakdown. A
dynamical mechanism like (hidden sector) gaugino condensation \cite{Nilles:1982ik,Nilles:1982xx,Ferrara:1982qs} can
explain a hierarchically small scale for the gravitino mass
\begin{align}
m_{3/2} \sim \frac{\Lambda^3}{M_\s{P}^2},
\end{align}
where $\Lambda$ denotes the renormalization group invariant scale of the
hidden sector gauge group and $M_\s{P}$ is the Planck mass. In a recent paper \cite{Derendinger:2005ed}, Derendinger, Kounnas and
Petropolous (DKP) identified a new solution in the framework of the heterotic
theory with an even stronger (quadratic) suppression of the gravitino mass
\begin{align}
m_{3/2} \sim \frac{\Lambda^6}{M_\s{P}^5}.
\end{align}
It was discovered in the study of moduli stabilization in the
$\mathbbm{Z}_2\times \mathbbm{Z}_2$
orbifold with NS-NS 3-form and geometric fluxes. While these
fluxes (combined with the gaugino condensate) are sufficient to stabilize
all moduli in many cases, they fail to do so in the new DKP-solution with
the double suppression of the gravitino mass. Two more moduli have to be
stabilized and it remains to be seen whether the doubly suppressed solution
survives in the process of moduli stabilization.

In the present paper, we try to study the new solution of DKP in a set-up
with all moduli fixed, define the criteria for the appearance of this
solution
and analyze its phenomenological properties. In the next chapter we shall
present
the model of DKP in its original form and notation. Chapter 3 addresses some
shortcomings of the model due to the presence of unfixed moduli. We conclude
that it is mandatory to have all moduli fixed before a meaningful analysis
of the
phenomenological properties can be performed. In
chapter 4  we list the basic
requirements for the appearance of DKP-like models and define a benchmark
model
where questions like the fine-tuning of the vacuum energy and the emerging
pattern
of the soft supersymmetry breaking term can be addressed. For this class of
models
we are again led to a kind of mirage pattern as previously
identified in \cite{Choi:2004sx,Choi:2005ge,Choi:2005uz,Endo:2005uy,Falkowski:2005ck,Baer:2006id}.
Chapter 5 is devoted to applications of this new solution. This includes
(most
importantly) scenarii with a TeV-scale gravitino and a condensation scale as
large
as a grand unified scale $M_\s{GUT}$ or the compactification scale
$M_\s{COMP}$. Other possible
applications include the generation of the $\mu$-term $(\mu H_1 H_2)$ in
models with a
small gravitino mass, as well as  cosmological applications concerning a
quintessential axion and the question of axionic
inflation.

\section{The model of DKP}
\label{genfra}

The analysis and the notation of this section will
follow \cite{Derendinger:2005ed}. We will start considering one important
phenomenological requirement, namely, moduli stabilization. To understand whether
moduli are stabilized or not we have to check the structure of the
scalar potential of the theory, so, as a first step, we dedicate
this section to the discussion of the possible terms which can
appear in the superpotential. We will consider two contributions:
fluxes and gaugino condensates.

\subsection{Fluxes, moduli and the superpotential}
\label{sec:flux}

We will confine our discussion to the $\mathbbm{Z}_2\times \mathbbm{Z}_2$
orbifold compactification of heterotic strings. This compactification
setup leads to seven main moduli (including the string dilaton) and
$N=1$ supersymmetry. The effective supergravity is the
$\mathbbm{Z}_2\times \mathbbm{Z}_2$ projection of the $N=4$
theory with sixteen supercharges. Extended supersymmetry is not
consistent with our phenomenological requirements,
since chiral fermions can exist in four
dimensions only when $N \leq 1$. For this reason it is mandatory
to reduce the amount of supersymmetry and the orbifold truncation
does this. Fluxes are introduced by gauging the $N=4$
supergravity \cite{Derendinger:2004jn,Derendinger:2005ph,Dall'Agata:2005ff,Andrianopoli:2005jv,Dall'Agata:2005mj,Dall'Agata:2005fm}: the parameters of the SUGRA theory
are the ``gauging structure constants'' and these are also the
flux parameters.

\subsubsection{Moduli identification}
\label{sub:moduli}

The identification of the moduli depends on the theory under
consideration. For heterotic strings on
${T^6/ \mathbbm{Z}_2\times \mathbbm{Z}_2}$, the moduli
are the dilaton-axion superfield $S$, the volume moduli
$T_A$ and the complex structure moduli $U_A$, $A=1,2,3$.
The index $A$ refers to the three complex planes left
invariant by $\mathbbm{Z}_2\times \mathbbm{Z}_2$. The
$N = 1$ supersymmetric complexification for these fields
is defined naturally in terms of the geometrical moduli
$G_{ij}$ (nine fields), the string dilaton $\Phi$ and
the components $B_{ij}$ (three fields) and $B_{\mu\nu}\sim a$
of the antisymmetric tensor. The indices $i\,,j$ run over the internal space
and $\mu\,,\nu$ are the Minkowski indices. Explicitly, the metric tensor restricted to the plane $A$ is
\begin{align}
\left( G_{ij} \right)_A &= \frac{t_A}{u_A} \left(\begin{array}{cc}
u_A^2 +\nu_A^2 & \nu_A \\
\nu_A  & 1 \
\end{array}\right) ,
\end{align}
with
\begin{align}
T_A = t_A+i\left( B_{ij}\right)_A,
\qquad\qquad  U_A= u_A + i \, \nu_A,
\end{align}
and
\begin{align}
e^{-2\Phi} = s(t_1t_2t_3)^{-1},  \qquad\qquad S=s+ ia.
\end{align}
The Weyl rescaling to the four dimensional Einstein frame is $G_{ij}=s^{-1} {\widetilde G}_{ij}$. The $\mathbbm{Z}_2\times \mathbbm{Z}_2$ projection of the $N=4$ theory leads to the scalar K\"ahler manifold
\begin{align}
M =& \left[\frac{SU(1,1)}{U(1)}\right]_S \times \prod^3_{A=1}\left[\frac{SU(1,1)}{U(1)}\right]_{T_A}\times \prod^3_{A=1}\left[\frac{SU(1,1)}{U(1)}\right]_{U_A},
\end{align}
with K\"ahler potential (in the usual string parameterization)
\begin{align}
K &= - \log \left(S+{\ov S}\right) -\sum_{A=1}^3 \log \left(T_A+{\ov T}_A\right)\left(U_A+{\ov U}_A\right).
\end{align}

It is our intention, on the one hand, to identify the components of the $B$-field which give the imaginary part of the $T$ moduli, on the other hand, to select the possible fluxes which can be ``turned on'' in this model. The first step is establishing which components of a $p$-form survive the $\mathbbm{Z}_2\times \mathbbm{Z}_2$ truncation. With this purpose in mind, let us write a generic $p$-form $\omega$ (for a review see \cite{Greene:1996cy}) as
\begin{align}
\omega &= \omega_{i_1,...,i_p} dx^{i_1} \wedge ... \wedge dx^{i_p},
\end{align}
where the coordinates $x^i$ are subject to the action of the orbifold as summarized in tab.\,\ref{tab:orbifold}. A component of a 3-form with one ``leg'' in each torus will certainly survive the truncation. This will reveal itself very useful in the next sections when dealing with the (combined) heterotic fluxes.

\begin{table}[t]
\small
\begin{center}
\begin{tabular}{w{2cm}v{1cm}v{1cm}v{1cm}v{1cm}v{1cm}v{1cm} }
\midrule\addlinespace
              & $x^5$ & $x^6$ & $x^7$ & $x^8$ & $x^9$ & $x^{10}$ \\\addlinespace\cmidrule(lr){1-1}\cmidrule(lr){2-7}
$\mathbbm{Z}_2$         & $-$ & $-$ & $-$ & $-$ & $+$ & $+$  \\
$\mathbbm{Z}^\prime_2$  & $+$ & $+$ & $-$ & $-$ & $-$ & $-$  \\\addlinespace
\midrule\midrule
\end{tabular}
\end{center}
\caption{Action of the $\mathbbm{Z}_2\times \mathbbm{Z}_2$ orbifold on the extra-dimensional coordinates.}
\label{tab:orbifold}
\end{table}

\subsubsection{Heterotic fluxes}
\label{sub:combined}

As first recognized in \cite{Derendinger:1985kk,Dine:1985rz,Strominger:1986uh,Rohm:1985jv}, possible fluxes in the heterotic theory are those of the modified NS-NS 3-form $\widetilde{H_3} = dB_2 + \ldots$, where the dots stand for the gauge and Lorenz Chern-Simons terms. There are eight independent real fluxes \cite{Derendinger:2004jn,Derendinger:2005ph}, invariant under the $\mathbbm{Z}_2 \times \mathbbm{Z}_2$ orbifold projection:
\begin{align}
\widetilde{H}_{579} \, , \;
\widetilde{H}_{679} \, , \; \widetilde{H}_{589} \, , \;
\widetilde{H}_{689} \, , \; \widetilde{H}_{5710} \, , \;
\widetilde{H}_{6710} \, , \; \widetilde{H}_{5810} \, , \;
\widetilde{H}_{6810} \, .
\end{align}
Leaving aside a systematic discussion, we just observe that, under the assumption of plane-interchange symmetry, there are four independent parameters (corresponding to gauging structure constants) which give $U$ dependent terms in the superpotential:
\begin{align}
\nonumber \widetilde{H}_{579} &\leftrightarrow 1, \\
\nonumber \widetilde{H}_{679} = \widetilde{H}_{589} = \widetilde{H}_{5710} &\leftrightarrow i \, (U_1 + U_2 + U_3), \\
\nonumber \widetilde{H}_{689} = \widetilde{H}_{5810} = \widetilde{H}_{6710} &\leftrightarrow - (U_1 \, U_2 + U_2 \, U_3 + U_1 \, U_3), \\
\widetilde{H}_{6810} &\leftrightarrow - i \, U_1 \, U_2 \, U_3.
\end{align}

The possible fluxes also include some geometrical ones \cite{Dall'Agata:2005ff,Andrianopoli:2005jv,Dall'Agata:2005mj,Dall'Agata:2005fm}, associated with the internal components of the spin connection $\omega_3$, and corresponding to coordinate dependent compactifications \cite{Scherk:1979zr}. These fluxes are characterized by real constants with one upper curved index and two lower antisymmetric curved indices
\begin{align}
f^i_{\;\; j k} &= - f^i_{\;\; k j} \, .
\end{align}
These constants must satisfy the Jacobi identities of a Lie group, $f^{i}_{\;\; j k} \ f^{k}_{\;\; l m} + f^{i}_{\;\; l k} \
f^{k}_{\;\; m j} + f^{i}_{\;\; m k} \ f^{k}_{\;\; j l} = 0$, and the additional consistency condition $f^{i}_{\;\; i k} = 0$
\cite{Derendinger:2004jn,Derendinger:2005ph}.

In agreement with the $\mathbbm{Z}_2 \times \mathbbm{Z}_2$ orbifold projection, we must assume here that
\begin{align}
f^{i_A}_{\quad i_B i_C} = 0 \qquad {\rm for} \;\;\;\; A=B \;\;
{\rm or} \;\; A=C \; {\rm or} \;\; B=C \, ,
\end{align}
which satisfies automatically the consistency condition $f^{i}_{\;\; i k}=0$. Geometrical fluxes are then described by 24 real parameters
\begin{align}
f^{i_A}_{\quad i_B i_C} \, , \qquad \big[(ABC)=(123),(231),(312)\big]\, ,
\end{align}
subject only to the Jacobi identities. The possible structures in the superpotential are $T$ dependent in the form
\begin{align}
\omega_{i_B i_C}^{i_A}  \leftrightarrow i T_A, T_A U_B,\, i T_A U_B U_C,\, T_A U_1 U_2 U_3.
\end{align}

Before proceeding with the analysis, one comment regarding the
compactification manifold is necessary. When we switch on fluxes, we are led to
heterotic theory on non-Calabi-Yau manifold. In a modern language
we can say that e.\,g. ``half-flat'' manifolds are
exploited implicitly in the DKP
model. The existence of these manifolds is strongly suggested by
type II mirror symmetry and, recently, heterotic theory on
half-flat has been discussed in \cite{Gurrieri:2004dt}, where a
Gukov-type formula has been obtained for the superpotential
induced by fluxes. In other words, the flux contribution to the
DKP-superpotential can be understood as a Gukov-type formula on a
half-flat manifold.

It is also important to recall that, in the heterotic theory,
$\widetilde{H}_3$ and $\omega_3$ fluxes can never generate an $S$-dependent
perturbative $N=1$ superpotential. Consequently, if our intention is to stabilize \emph{all} moduli,
including the dilaton, some additional stabilizing contribution
must be included. For this reason we will therefore consider
nonperturbative effects, in particular, gaugino condensation. For
further details on fluxes in heterotic theory the reader is
referred to the literature
\cite{Gurrieri:2007jg,Gurrieri:2004dt,Brustein:2004xn,Becker:2003yv,LopesCardoso:2003sp}.

\subsection{Supersymmetry breaking in Minkowski space}
\label{sub:minkowski}

In a general supergravity theory with K\"ahler potential $K = -\sum_j\log( Z_j + \ov Z_j)$ where $Z_j$ denotes the moduli, supersymmetry is spontaneously broken if the equations
\begin{align}
F^j  \equiv W - \klammer{Z_j + \ov Z_j} W_j =0
\end{align}
cannot be solved for all scalar fields $Z_j$ (and with $\real Z_j>0$). DKP make a specific ansatz demanding broken supersymmetry in Minkowski space, which implies
\begin{align}
\langle V \rangle = 0 \, , \qquad \langle W \rangle \ne 0 \,.
\label{eqn:dkpminkowski}
\end{align}
A stationary point is found from $\partial_j V =0$, $\forall j$, which explicitly reads
\begin{align}
0 = e^{-K}V K_j  - \ov W_j F^j - 3  W_j\ov W + \sum_{i\ne j} \left[ W_j - \klammer{Z_i+\ov Z_i}W_{ij} \right] \ov F^i - \klammer{Z_j + \ov Z_j} W_{jj} \ov F^j\, ,
\label{eqn:dkpstationary1}
\end{align}
for each scalar field $Z_j$. The first term vanishes in Minkowski space and the second derivative of the superpotential $W_{jj}$ is nonzero only for the moduli appearing in the exponential gaugino condensate.

The requirements eq.\,(\ref{eqn:dkpminkowski}) split the scalar
fields into two categories, with either $\langle W_j \rangle=0$
and $\langle F^j \rangle \ne 0$ or $\langle F^j \rangle =0$. Only
the first category is involved in SUSY breaking.
Tab.\,\ref{tab:cathegories} shows the field content of each
category. Taking this partition of the fields into account, the
stationary point condition eq.\,(\ref{eqn:dkpstationary1}) breaks
down to seven conditions
\begin{align}
0 = \sum_{i\ne j} \langle W_{ij}\real Z_i \rangle ,
\label{eqn:dkpstationary2}
\end{align}
with summation restricted over moduli which break supersymmetry.

\begin{table}[t]
\small
\begin{center}
\begin{tabular}{w{3cm}v{3cm}v{3cm}}
\midrule\addlinespace
& $\left\langle W_j \right\rangle=0$ & $\left\langle F_j \right\rangle=0$ \\\addlinespace\cmidrule(lr){1-1}\cmidrule(lr){2-3}
Fields         & $T_1$, $T_2$, $T_3$ & $S$, $U$  \\
\cancel{SUSY}  & $+$ & $-$  \\\addlinespace
\midrule\midrule
\end{tabular}
\end{center}
\caption{Categories of fields in the DKP model.}
\label{tab:cathegories}
\end{table}

\subsection{The double suppression of the gravitino mass}
\label{sub:dkp}

Consider for concreteness a superpotential of the type \cite{Derendinger:2005ed}
\begin{align}
W &= 3 \widehat{A} U +  \widehat{D} U_4,
\label{eqn:dkpsuper}
\end{align}
where $U_4 = U^3$. We have introduced the following functions of $T_1,T_2$ and $S$:
\begin{align}
\widehat{A} &= \Big[\alpha + \alpha^\prime w(\widetilde{S})\Big]\xi + Aw(\widetilde{S}),\\
\widehat{D} &= \Big[\delta + \delta^\prime w(\widetilde{S})\Big]\xi +
Dw(\widetilde{S}),
\end{align}
with $\xi = T_1-T_2$ and $w(\widetilde{S})=\mu^3 e^{-\widetilde{S}}= \Lambda^3$ where $\mu \simeq M_\s{P}$ is the
RG scale\footnote{The RG scale $\mu=\abs{\mu}\exp{(i\, \phi_{\mu})}$ can be consistently taken real by shifting the
imaginary part of the dilaton in the heterotic theory. In this work we
will assume $\phi_{\mu}=0$.}, $\Lambda$ is the RG invariant
scale of the confining gauge group and
\begin{align}
\frac{24 \pi^2 S}{b_0} &\longrightarrow \widetilde{S}
\label{eqn:rescaling}
\end{align}
where $b_0$ is the one-loop beta function coefficient.

The minimization condition (\ref{eqn:dkpstationary2}) reads $\real
\xi =0$. We will therefore choose $\widetilde{S}=\widetilde{s}
-i{\pi/2}$ and $U= u$ real. Everything is consistent provided
$\alpha,\delta$ and $A,D$ are real and
$\alpha^\prime,\delta^\prime$ are imaginary.

The no-scale requirement $\langle V \rangle =0$ is fulfilled provided $\langle W_{T_1} \rangle = \langle W _{T_2} \rangle= 0$ ($W$ is independent of $T_3$) and $F^S=F^U=0$. This allows us to express $\xi$ and $u$ as functions of $s$. In particular
\begin{align}
u(\widetilde{s}) &= \sqrt{\frac{\widehat A}{\widehat D}}, \label{eqn:uofs}\\
\xi(\widetilde{s}) &= -\frac{1}{4}
\frac{3D\alpha+A\delta+\klammer{3D\alpha^\prime+A\delta^\prime}w}{\klammer{\alpha+\alpha^\prime
w}\klammer{\delta+\delta^\prime w}} w.\label{eqn:ksiofs}
\end{align}
The central equation for the determination of $\widetilde{s}$ is
given by:
\begin{align}
\frac{~2~}{\widetilde{s}} &= -4 - \frac{\klammer{\alpha^\prime
\delta-\delta^\prime \alpha}w}{\klammer{\alpha+\alpha^\prime
w}\klammer{\delta+\delta^\prime w}} \frac{3D\alpha+A\delta +
\klammer{3\alpha^\prime
D+A\delta^\prime}w}{D\alpha-A\delta+\klammer{D\alpha^\prime-A\delta^\prime}w}.
\label{eqn:sdet}
\end{align}

As previously discussed in \cite{Derendinger:2005ed}, eq.\,(\ref{eqn:sdet})
admits physically acceptable solutions for $\widetilde{s}$,
provided that the fluxes $\alpha,\delta, \alpha^\prime,
\delta^\prime,A,D$ are large while their ratios are of order
unity. If this requirement is fulfilled we can define a variable
$\rho$ (real function of $\widetilde{s}$) as
\begin{align}
\rho =i\frac{D\alpha-A\delta}{D\alpha w},
\label{eqn:dkprho}
\end{align}
which can be consistently taken to be of order one since $w$ is
small and $D\alpha/A\delta$ is of order one. As we shall discuss later,
this requires a certain amount of fine tuning for
$D\alpha - A\delta \ll 1$.

The gravitino mass is given by \cite{Derendinger:2005ed}:
\begin{align}
\nonumber e^{-K/2} m_{3/2} &= \langle W\rangle \\
&= \frac{A\delta^\prime - D\alpha^\prime + \frac{A \delta - D\alpha
}{w}}{\alpha+\alpha^\prime w}\left(-3\frac{\alpha+\alpha^\prime w}{\delta+\delta^\prime w}\right)^{3/2}w^2\, ,
\end{align}
for generic plane symmetric situations. In the special case $\alpha^\prime=i \alpha$ and $\delta^\prime=-i \delta$ and within the above approximations, the result is
\begin{align}
\nonumber e^{-K/2}m_{3/2} & \approx i \left( 2D e^{-2 \widetilde{s}} + \frac{A \delta -D \alpha}{\alpha} e^{-\widetilde{s}}\right) \\
& \approx  i4D\left(-
\frac{3\alpha}{\delta}\right)^{3/2}\frac{\widetilde{s}}{2\widetilde{s}+1}
w^2. \label{eqn:dkpgravitino}
\end{align}
The gravitino mass scales as $w^2$ and this is due to the absence of flux-induced $S$-term in the heterotic superpotential.

It is relevant to discuss which values of the parameters will give
a small $w$. If we consider a Planckian scale $\mu$ then
$\widetilde{s} \sim 10$ is necessary to have a small $w$ and to
achieve a reasonable value of $g_\s{GUT}$.
Eq.\,(\ref{eqn:dkpgravitino}) written in the form shows that the
$S$ modulus is stabilized through the presence of the condensate.
Strictly speaking, this is not a ``racetrack'' mechanism proper
\cite{Krasnikov:1987jj} because we only have one condensate. However, the
condensate enters into the superpotential in a rather complex way
and several terms are added together, so, as far as our model is
concerned, it gives a result that otherwise can only be achieved
by a racetrack mechanism.

\section{Problematic aspects of the model}

Up to now the stabilization of the moduli can be summarized as follows:
\begin{itemize}
\item $S$ can be stabilized at an acceptable value,
\item $U$ is real and stabilized through fluxes,
\item $\xi=T_1-T_2$ is  fixed as a function of $S$ and
$\real\xi=0$ from eq.\,(\ref{eqn:dkpstationary2}),
\item $T_3$ and $T_1+T_2$ are flat directions and thus not stabilized.
\end{itemize}

\subsection{Unfixed moduli}

The problematic aspect of the DKP-strategy is the appearance of
unfixed $T$ moduli. It is a consequence of the restricted no-scale
ansatz eq.\ (\ref{eqn:dkpminkowski}) and it leads to a situation
with $F^S=0$ and $F^T \neq 0$. This ansatz requires $E_\s{VAC}=0$
in a situation where not all moduli are fixed. It remains to
be seen whether this strategy is the most appropriate one since
the goal of the procedure is a vanishing $E_\s{VAC}$ after
fixing all moduli.

One way to proceed is the application of the old local no-scale
idea, where one assumes corrections to the K\"ahler potential which fix the
remaining moduli while keeping $E_\s{VAC} = 0$.
Thus to remove the flat directions we will modify the K\"ahler potential
demanding the flatness condition ($V\equiv0$) only \emph{locally} \cite{Cremmer:1983bf,Ellis:1984bm}.
In the remaining part of this section we will focus on this
point (following the notation in \cite{Cremmer:1983bf}).
The final outcome will be the stabilization of all moduli.

\begin{figure}
\captionsetup[figure]{labelfont={footnotesize,bf},textfont=footnotesize,labelsep=mysep,labelformat=mypiccap,format=default,justification=RaggedRight,width=7cm,indent=5pt}
\begin{minipage}{0.5\linewidth}
\includegraphics[width=\linewidth]{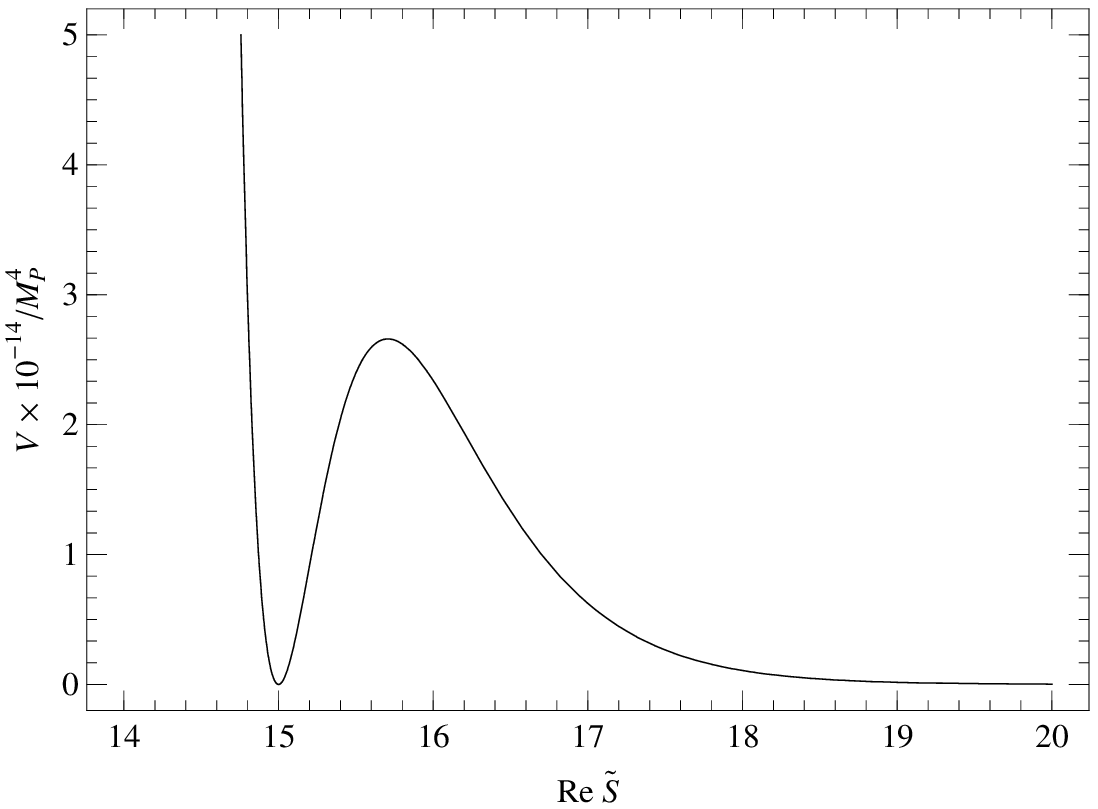}
\caption{Scalar potential for the dilaton.}
\label{fig:sdir}
\end{minipage}
\begin{minipage}{0.5\linewidth}
\includegraphics[width=\linewidth]{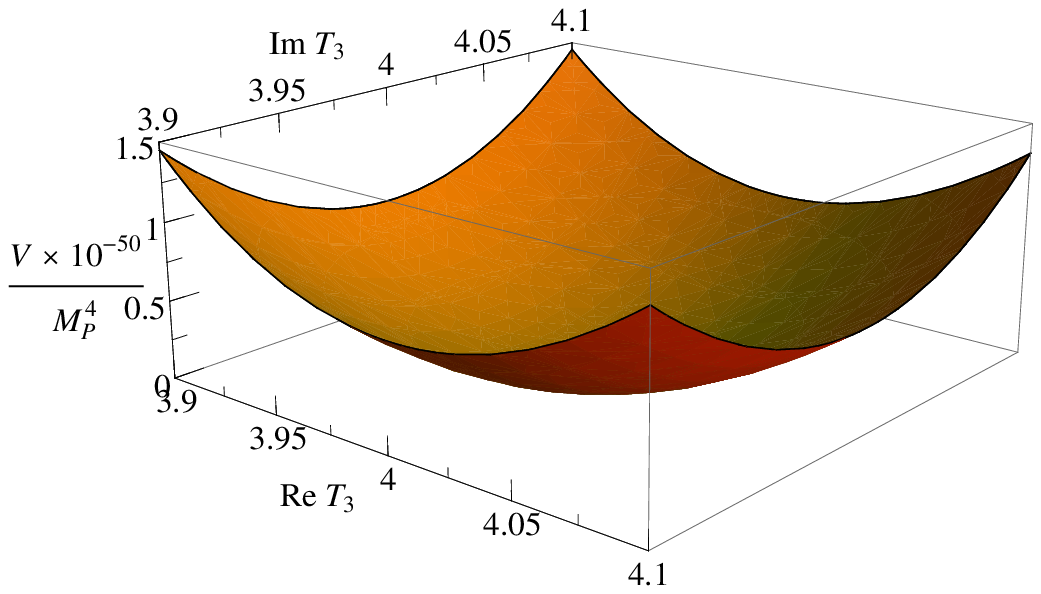}
\caption{Stabilizing potential for $T_3$ displayed in the complex plane.}
\label{fig:tdir}
\end{minipage}
\end{figure}

For the flatness condition to be around point $z_0$ in $\cal D$ (the positive kinetic energy domain $G_{z\ol{z}}>0$), we demand
\begin{align}
\partial_z \partial_{\ol{z}} e^{-G/3} = \phi_{z\ol{z}} (z,\ol{z}),
\label{eqn:ode}
\end{align}
where the real function $\phi(z,\ol{z})$ satisfies the conditions
\begin{align}
\phi_{z\ol{z}}\geq 0, \forall z \in \cal{D}
\end{align}
and
\begin{align}
\phi_{z\ol{z}}(z_0,\ol{z}_0)=0.
\end{align}
The general solution of eq.\,(\ref{eqn:ode}) is
\begin{align}
G=-\frac{3}{2}\log\klammer{f+\ol{f} + \phi}^2,
\label{stabilizingphi}
\end{align}
with the positive kinetic energy domain defined by
\begin{align}
G_{z\ol{z}}=3\frac{\abs{f_z+ \phi_z}^2-\phi_{z\ol{z}} \left(f+\ol{f} +
\phi\right)}{\left(f+\ol{f} + \phi\right)^2}>0.
\label{eqn:positivity}
\end{align}
The corresponding scalar potential is positive definite in $\cal{D}$, provided that $\phi_{z\ol{z}}\geq0$ as well as $f+\ol{f} + \phi >0$ as can be seen from its analytic expression
\begin{align}\label{localnoscalepot}
V &= 3 \frac{\phi_{z\ol{z}}\left(f+\ol{f} + \phi\right)}{\abs{f+\ol{f} + \phi}^3
\left[\abs{f_z + \phi_z}^2 - \phi_{z\ol{z}} \left(f+\ol{f} + \phi\right)\right]}.
\end{align}
The K\"ahler correction is designed in such a way that it vanishes at $z_0$. Thus, the presence of $\phi$ in the scalar potential deforms its shape only outside $z_0$. Let us for concreteness consider
\begin{align}
\phi &= \frac{\klammer{z-z_0}^2\klammer{\ol{z}-\ol{z}_0}^2}{4},
\label{eqn:kahlercorrection}
\end{align}
and apply the analysis to $z=T_3$. As evident from eq.\,(\ref{eqn:kahlercorrection}) we obtain $\phi_{z\ol{z}}=\left|T_3-z_0\right|^2$ and the positivity condition eq.\,(\ref{eqn:positivity}) is fulfilled. The scalar potential exhibits a local minimum at $z_0$. Fig.\,\ref{fig:tdir} illustrates the scalar potential in the complex $T_3$ plane. A similar procedure can be exploited to remove the $T_1+T_2$ flat direction. Note that due to the K\"ahler correction $\phi$ the K\"ahler manifold no longer exhibits the $SU(1,1)$ symmetry. In the minimum $z_0$, however, the stabilization of $S$ does not clash with the stabilization of $T_3$ and $T_1+T_2$ since in $z_0$ the correction $\phi$ and its derivatives vanish.

Since we have
stabilized now all the moduli in the presence of broken supersymmetry,
we can start to analyze explicitely
the resulting pattern of soft supersymmetry breaking terms.

\subsection{The pattern of supersymmetry breakdown}

\subsubsection[Tree-level $F$-terms and $m_{3/2}$]{Tree-level $\boldsymbol{F}$-terms and $\boldsymbol{m_{3/2}}$}

To obtain a phenomenologically attractive gravitino mass we shift
the position of the minimum in the $S$ direction to
$\widetilde{S}_0=15-i\pi/2$. This corresponds, by reversing the
rescaling eq.\,(\ref{eqn:rescaling}), to the gauge group $SU(11)$.
We choose the minima in the $T$ directions to be at $T_0=4+i4$. A
Minkowski vacuum is obtained e.\,g. for the following set of flux
coefficients
\begin{align}
\alpha=100,\, \delta=-100,\, A=10,\, D=-10.00001.
\end{align}
This choice respects the condition that eq.\,(\ref{eqn:dkprho}) is of
order one and we have a doubly suppressed gravitino mass. In greater
detail, numerically $(A \delta-D \alpha)/(\alpha w) \simeq 33$ and
we keep under control the dangerous $1/w$ contribution, that could
spoil the double suppression in the gravitino mass formula,
granted that we accept a mild fine-tuning\footnote{It is worthwhile
to recall that this fine-tuning is less severe than the KKLT one
which is of order $10^{-16}$ \cite{Kachru:2003aw}.} of the parameters of order $A
\delta-D \alpha \simeq 10^{-3}$.

In this particular vacuum we obtain
\begin{align}
F^S &= F^U=0, \\
F^{T_1} &= F^{T_2} = F^{T_3} = -8 m_{3/2},
\end{align}
where the gravitino mass is given by $m_{3/2}=1.11\times 10^{-14}M_\s{P}$. Remarkably the local no-scale structure exploited to stabilize the flat directions does not modify the result $F^S=0$. At first sight, the result $F^{T_1}=F^{T_2}=F^{T_3}$ might be surprising since the $T$ moduli have been considered in a strongly asymmetric way in the superpotential and during the process of stabilization. However, we must remember that $W_{T_1}=W_{T_2}=W_{T_3}=0$ was one of the conditions to obtain a Minkowski (no-scale) solution and, once exploited in the evaluation of the $F$ terms, it will give us the very symmetric configuration $F^{T_1}=F^{T_2}=F^{T_3}$.

\subsubsection{Tree-level soft terms}

For the evaluation of the soft supersymmetry breaking terms we include the contribution of matter fields in the K\"ahler potential. We will focus our attention on $(0,2)$ symmetric orbifolds. If we denote collectively the $T$ and the $U$ moduli by $T_p$, where e.\,g. $T_p=U_{p-3}$ with $p=\{3,4,5\}$, the K\"ahler potential becomes \cite{Brignole:1995fb,Ibanez:1992hc}
\begin{align}
K &= -\log\klammer{S+\ol{S}}- \sum_p \log\klammer{T_p+\ol{T}_p} + \sum_\alpha Q_\alpha
\ol{Q}_\alpha \prod_p \klammer{T_p + \ol{T}_p}^{-n^p_\alpha},
\end{align}
with $Q_\alpha$ being the observable matter fields and $n^p_\alpha$ are the so-called \emph{modular weights}. As already mentioned before, in our vacuum we have $F^S=F^U=0$. Consequently the structure of the K\"ahler potential simplifies to
\begin{align}
K &= - \sum_{i=1}^3 \log\klammer{T_i+\ol{T}_i} + \sum_\alpha Q_\alpha \ol{Q}_\alpha \klammer{T_1 +\ol{T}_1}^{-n^1_\alpha} \klammer{T_2 +\ol{T}_2}^{-n^2_\alpha} \klammer{T_3 + \ol{T}_3}^{-n^3_\alpha}. \label{Kpot}
\end{align}
We evaluate scalar masses, soft trilinear couplings and gaugino masses from (see e.\,g. \cite{Brignole:1993dj,Brignole:1997dp})
\begin{align}
m^2_\alpha &= m_{3/2}^2 - \ol{F}^{m} F^n \partial_m \partial_n \log\mathcal{K_\alpha} ,\\
A_{\alpha \beta \gamma} &= F^m \left[\widehat{K}_m + \partial_m \log Y_{\alpha \beta \gamma}- \partial_m
\log\klammer{\mathcal{K_\alpha} \mathcal{ K_\beta} \mathcal{ K_\gamma}}\right] ,\\
M_{a} &= \frac{1}{2} \klammer{\real f_a}^{-1} F^m \partial_m f_a,
\end{align}
where $m$ runs over SUSY breaking fields, $\widehat{K}$ is the K\"ahler potential for the hidden sector fields, $\mathcal{K_\alpha}=\frac{\partial^2 K}{\partial Q_\alpha \partial \ol{Q}_\alpha}$ and we will assume that $Y_{\alpha\beta\gamma}$ are moduli independent. We obtain
\begin{align}
m_\alpha^2 &= m_{3/2}^2\big[1- n^1_\alpha-n^2_\alpha-n^3_\alpha\big], \\
A_{\alpha\beta\gamma} &= m_{3/2} \Big[\klammer{1-n^1_\alpha-n^1_\beta - n^1_\gamma}+
\klammer{1-n^2_\alpha-n^2_\beta - n^2_\gamma}+\klammer{1-n^3_\alpha-n^3_\beta -
n^3_\gamma}\Big], \\
M_{a} &= 0,
\end{align}
where $n_\alpha^1+n_\alpha^2+ n_\alpha^3=1$. At tree-level we have really a no-scale configuration for the soft terms: the scalar masses, the trilinear couplings and the gaugino masses are vanishing.

\subsubsection{Quantum corrections}

In the previous section we obtained vanishing \emph{tree-level}
soft terms. To construct a realistic model we can take into
account quantum effects like anomaly mediation
\cite{Randall:1998uk} and/or threshold corrections to the gauge
kinetic function \cite{Dixon:1989fj,Dixon:1990pc,Derendinger:1991hq,Derendinger:1991kr}. However, since loop corrections
exceedingly complicate the evaluation of the soft terms, we will
dedicate the next section entirely to the discussion of a simple
benchmark model encompassing the main features of the DKP set-up. In
this section instead, we will summarize the modifications induced
by threshold corrections in our model (in the gauge kinetic
function, in the superpotential and in the K\"ahler potential).

The detailed structure of quantum corrections depends on the model we consider (the case of $\mathbbm{Z}_2 \times \mathbbm{Z}_2$ orbifold model is discussed in \cite{Petropoulos:1996rr,Bailin:1994ma,Nilles:1997vk}) and usually these corrections are complicated functions of the moduli \cite{Dixon:1989fj,Dixon:1990pc,Derendinger:1991hq,Derendinger:1991kr}. However for our purposes it will be sufficient to consider the following modification to the gauge kinetic
function (see \cite{Ibanez:1986xy})
\begin{align}
f=S+ \epsilon T,
\label{eqn:oneloopf}
\end{align}
where the explicit $T$ dependence will lead us to nonvanishing gaugino masses and $\epsilon$ is a small number that we now specify. The one-loop gauge coupling constant is given by the real part of a gauge kinetic function of the form \cite{Lust:1991yi}
\begin{align}
f_\s{1-LOOP}=S-\frac{1}{8 \pi^2} \sum_i \klammer{\frac{b_0}{3}-\delta_\s{GS}^i}\log\eta^2(T_i),
\end{align}
where $b_0$ is a beta function coefficient and the mixing with coefficient $\delta_\s{GS}^i$ generalizes the Green-Schwarz mechanism. For
large $T$ we have $\log\eta(iT) \sim - \frac{\pi}{12} T$, and in our $\mathbbm{Z}_2 \times \mathbbm{Z}_2$ orbifold model $\delta_\s{GS}^i=0$ \cite{deCarlos:1992da}. In this way we recover the simpler structure of eq.\,(\ref{eqn:oneloopf}).

The superpotential of the model contains the
\emph{nonperturbative} factor $w(\widetilde{S})= e^{-\widetilde{S}}$ which
depends crucially on the value of the gauge coupling constant $g$.
At loop-level $g$ is not specified {\it only} by the dilaton field
$S$, but also by the $T$ field. For this reason we will substitute
the $S$ field in the superpotential with the 1-loop corrected
expression $S+ \epsilon T$ (see above).

In case of a pure Yang-Mills theory, the one-loop K\"ahler potential is given by \cite{Derendinger:1991hq, Lust:1991yi, Nilles:1997vk}
\begin{align}
K_\s{1-LOOP} =-\log\left[ \klammer{S+\ov S} -\frac{1}{8\pi^2}\sum_{i=1}^3 \delta_\s{GS}^i\log\klammer{T_i+\ov T_i} \right] -\sum_{i=1}^3\log\klammer{T_i+\ov T_i}.
\end{align}
$K_\s{1-LOOP}$ now leads to a mixing between the $S$ and the $T_i$ fields, but in our orbifold model $\delta_\s{GS}^i$ is vanishing and no correction will be present \cite{de Carlos:1992da}.

This mixing of $S$ and $T$ will induce soft supersymmetry breaking
terms in the observable sector, including nontrivial gaugino masses.
Of course, the mixing will also induce a shift in the location of
the minimum of the effective potential. Such a shift will generically
lead to a nonvanishing value of the vacuum energy as well. A further
fine tuning is required to obtain a suitable value for
$E_\s{VAC}$. A complete treatment
of the effective potential is rather complicated and will not be discussed
here in detail. We shall rather adopt a strategy different from DKP
in order to simplify the discussion.

\subsection{Intermediate conclusions}

As we have seen, we can fix the remaining moduli and keep vanishing
vacuum energy. The strategy adopted so far, however, has several drawbacks.
The corrections to
the K\"ahler potential in eq.\,(\ref{stabilizingphi})
are chosen ad hoc and there is no
 convincing theoretical justification (e.\,g. from string theory
considerations). The procedure is such that $F^S=0$ is frozen in
and we have problems with the soft supersymmetry breaking terms,
e.\,g. vanishing tree level gaugino masses. It is also well known
that radiative corrections spoil the no-scale structure.
As we have seen above, generically these
corrections induce a mixing of $S$ and $T$ and therefore $F^S \neq 0$
as well as a nonvanishing contribution to $E_\s{VAC}$. Thus the
vacuum energy has to be tuned a second time. We therefore conclude
that the no-scale strategy of DKP is not necessarily the most
appropriate one. One should first fix all moduli (without
intermediate assumptions on $E_\s{VAC}$) and then care about
(the tuning of) the vacuum energy once and for all.

This is what we are trying to do in the next chapter. The question
is whether the observed double suppression of the gaugino
condensate can be realized in this more general set-up as well, or
whether it is tied to the specific no-scale ansatz eq.\,(\ref{eqn:dkpminkowski}) adopted by DKP.

\section{Towards a resolution of the problems}

\subsection{Basic ingredients}

Let us recapitulate the basic ingredients needed for the double
suppression. The obvious requirement is the absence of $S$ in the
perturbative superpotential (only nonperturbative contributions
of the form $e^{-S}$ are present) which is automatically fulfilled in the heterotic
theory. We also need some tuning of parameters as explained in
the last section ($A \delta - D\alpha \ll 1$). One should also note
that terms with $e^{-S}$ in the superpotential need to be multiplied
by non-trivial functions of the $T$ moduli (a generic result in
the heterotic theory originating in world sheet instanton
effects \cite{Hamidi:1986vh,Dixon:1986qv}). Last but not least we need a superpotential with terms
that allow large masses for the $T$ moduli, although the
classical superpotential does not include quadratic terms in $T$ (but
only constant and linear terms). This requirement has been
discussed in detail in \cite{Choi:2004sx} both for the heterotic and the type
IIB case, and it strongly relies on the existence of the
complex structure moduli. Once these have been fixed and integrated
out we are left with an effective superpotential $W_\s{EFF}(S,T)$
which could include terms quadratic (and higher order) in $T$. Of course,
one could also consider more general compactification than
the $\mathbbm{Z}_2 \times \mathbbm{Z}_2$ orbifolds considered by DKP
that allow for more general superpotentials.
In any case, the above requirements seem to be quite easily
fulfilled in the framework of heterotic
string theory\footnote{It would be interesting to see whether such
a situation could also appear in the framework of type IIB theory.}.

We thus consider the heterotic theory with $S$, $T$, and $U$ moduli
as given in DKP. In a first step we use fluxes to fix the $U$ moduli
without breaking supersymmetry. The $U$ moduli become heavy and
can be integrated out leading to an effective superpotential
$W_\s{EFF} (S,T)$
where $S$ only appears through the condensate $e^{-S}$. The
simplest form to realize the doubly suppressed solution of DKP then reads:
$W_\s{EFF} \sim T e^{-S} + T^2$
where we dropped numerical  coefficients for the moment.
The equation of motion for $T$ then relates $T \sim e^{-S}$ and
$\left\langle W_\s{EFF}\right\rangle \sim (e^{-S})^2$ ,
as desired. The mild fine-tuning of DKP ($A \delta - D\alpha \ll 1$)
has a counterpart in our benchmark model: the coefficient of a possible term
linear in $T$ has to be small, otherwise the double suppression would
be spoiled. Thus this simple model captures all the aspects of the
DKP-model with one exception: the $S$ modulus is not yet fixed.
In fact, we are faced with a potential that shows run-away behavior for
$S \rightarrow +\infty$, and the potential is positive. Therefore we have
to find a mechanism to fix $S$ and tune the vacuum energy
to an acceptable value.
Both problems can be solved simultaneously by adopting
the ``downlifting strategy'' explained in \cite{Lowen:2008fm}. This will be
explicitely worked out in the remainder of this section.

\subsection{A benchmark model}
\label{sub:benchmark}

Given the guidelines in the previous section the purpose of this part
is to construct and analyze a simpler framework covering the key
features of the DKP model. The fact that the only $S$-dependence
of the superpotential in the DKP model is encoded in the gaugino
condensate, can be identified as the crucial requirement for the double
suppression. We would like to express the complicated form of the
superpotential eq.\,(\ref{eqn:dkpsuper}) in a more transparent language.
In what follows we consider $S$, $T$ and $U$ moduli. After the $U$
moduli have been integrated out we assume the effective superpotential
to be of the form
\begin{align}
W_\s{EFF} &= \mathcal{A}_0 e^{-aS} T + \mathcal{A}_1 T + \mathcal{A}_2 T^2 + \cdots + \mathcal{A}_n T^n ,
\label{eqn:effsuper}
\end{align}
where $\mathcal{A}_0$,\dots,$\mathcal{A}_n$ and $a$ are real constants.
For the case of simplicity we are considering a real dilaton field $S$
and one single real K\"ahler modulus $T$. We fine-tune the coefficient
$\mathcal{A}_1$ to be small (see discussion above) such that
the term linear in $T$
becomes negligible.

The equation of motion for the $T$ modulus $F^T=0$ reads
\begin{align}
0 &= W\del_T K + \del_T W \\
&= \mathcal{A}_0\klammer{-\frac{3}{2}+1}e^{-aS} + \mathcal{A}_2\klammer{-\frac{3}{2}+2}T + \cdots + \mathcal{A}_n\klammer{-\frac{3}{2}+n}T^{n-1}.
\label{eqn:eom}
\end{align}
For eq.\,(\ref{eqn:eom}) to be satisfied, $T$ has to be small which implies that $T^3$ and higher powers of $T$ can be safely neglected
in eq.\,(\ref{eqn:effsuper}). From the equation of motion eq.\,(\ref{eqn:eom}) one obtains
\begin{align}
T &= \frac{\mathcal{A}_0}{\mathcal{A}_2}e^{-aS}.
\end{align}
Consequently we can integrate out the $T$ field and end up with the effective superpotential
\begin{align}
W_\s{EFF} &= 2\frac{\mathcal{A}_0^2}{\mathcal{A}_2} e^{-2aS}.
\label{eqn:effsuper2}
\end{align}
This is exactly the double suppression as obtained in the DKP model.
At this stage, however, the dilaton is not yet stabilized since the scalar
potential from eq.\,(\ref{eqn:effsuper2}) shows a run-away behavior.
The remaining
task to perform is to stabilize the dilaton and assure a reasonable
vacuum energy. As was recently shown \cite{Lowen:2008fm} these two operations can be
done economically in one step.

Following the discussion of \cite{Lowen:2008fm} we consider the impact of hidden sector matter through the interaction with the effective theory obtained after integrating out $U$ and $T$ moduli. For concreteness and simplicity we will focus on a Polonyi-type superpotential \cite{Polonyi:1977pj} so that the full superpotential is given by
\begin{align}
\nonumber W &= W_\s{EFF}(S) + W_\s{POLONYI}(C) \\
&= -A e^{-2aS} + \nu + \tau^2 C,
\label{eqn:fullsuper}
\end{align}
where we have chosen $\mathcal{A}_0=\mathcal{A}_2=-A/2$, $\nu$ and $\tau^2$ are constants and $C$ represents a hidden sector matter field, assumed to be a gauge singlet. The corresponding K\"ahler potential is
\begin{align}
K &= -\log{\klammer{S+\ol{S}}} + C\ol{C}.
\label{eqn:newkahler}
\end{align}

As shown in \cite{Lowen:2008fm}, systems of this type are capable of changing the shape of the run-away scalar potential and lead to formation of stationary points. The stationary point in the configuration eq.\,(\ref{eqn:fullsuper},\,\ref{eqn:newkahler}) turns out to be a local minimum. By appropriately choosing the parameters of the Polonyi subsector the cosmological constant can be adjusted/finetuned to the desired value.

\begin{figure}
\captionsetup[figure]{labelfont={footnotesize,bf},textfont=footnotesize,labelsep=mysep,labelformat=mypiccap,format=default,justification=RaggedRight,width=7cm,indent=5pt}
\begin{minipage}{0.5\linewidth}
\includegraphics[width=\linewidth]{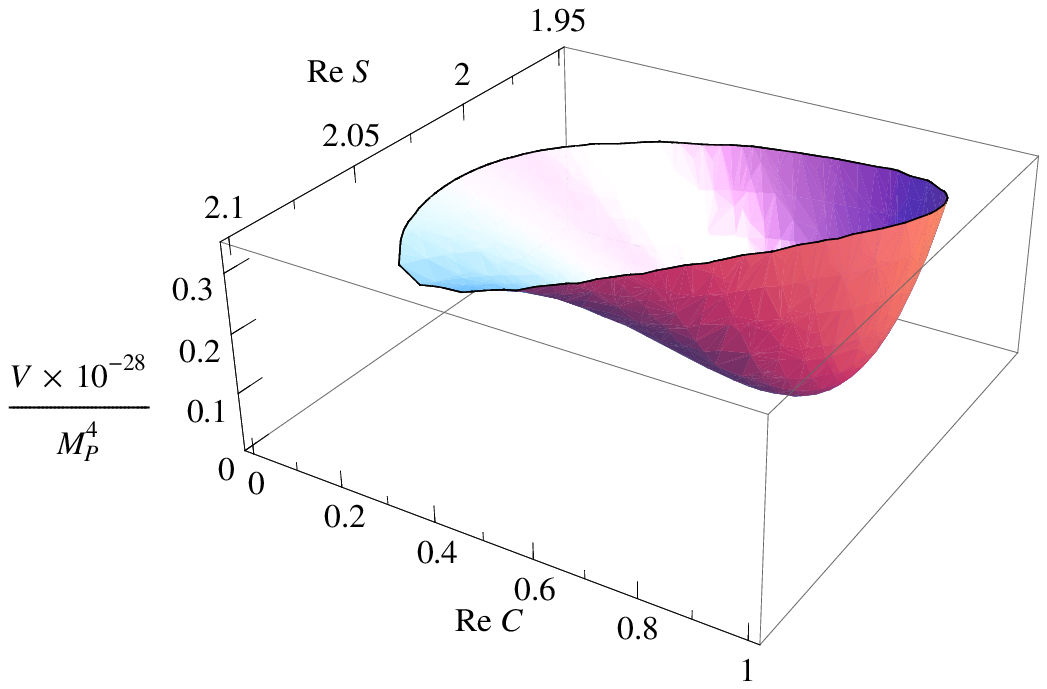}
\caption{Scalar potential for the $S$-$C$ system.}
\label{fig:benchmark}
\end{minipage}
\begin{minipage}{0.5\linewidth}
\includegraphics[width=\linewidth]{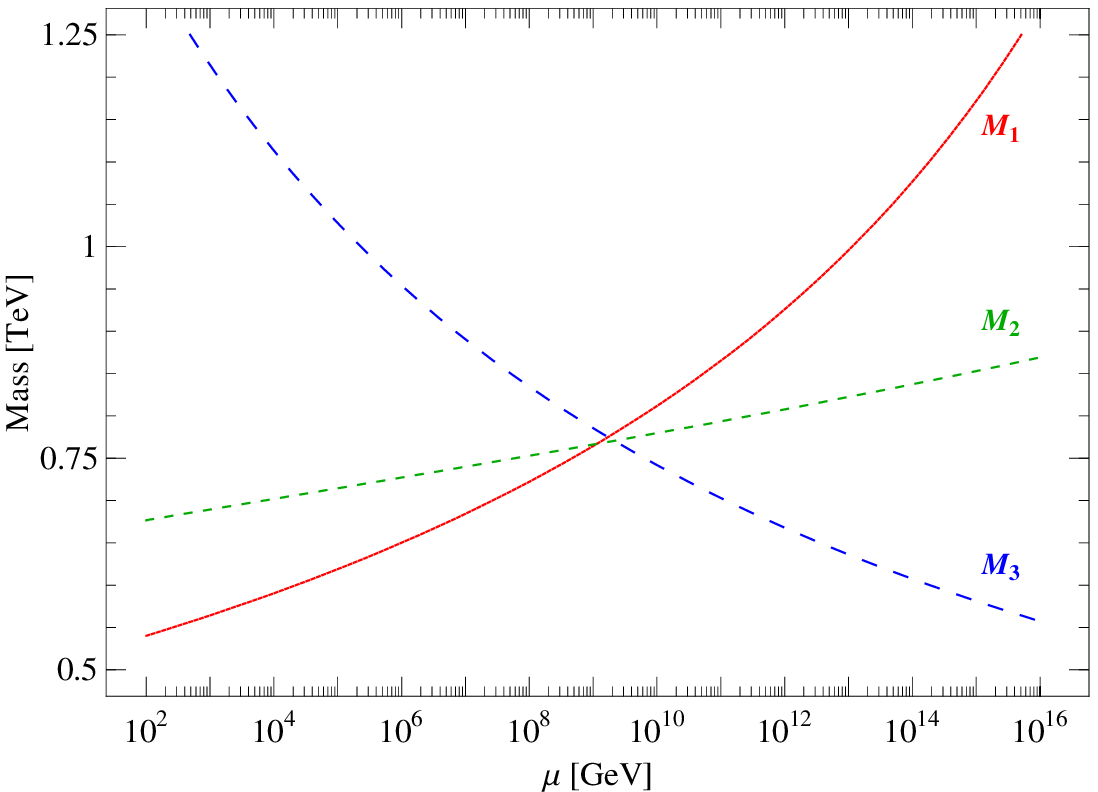}
\caption{RG evolution of the gaugino masses.}
\label{fig:gauginos}
\end{minipage}
\end{figure}

The consequence of the $F$-term uplifting/downlifting is the appearance
of a so-called \emph{little hierarchy} \cite{LoaizaBrito:2005fa,Lebedev:2006qq,Lowen:2008fm,GomezReino:2006dk} originating from the factor
\begin{align}
\Upsilon = \log{\klammer{\frac{M_\s{P}}{m_{3/2}}}} \sim \mathcal{O}\klammer{4\pi^2} .
\end{align}
In particular it leads to a suppression of the dilaton contribution to the soft terms
\begin{align}
F^S &\sim \frac{m_{3/2}}{\Upsilon},
\label{eqn:fs}
\end{align}
such that SUSY breaking is dictated by the matter sector, since $F^C\sim m_{3/2}$. The scale of the soft terms is set by the gravitino mass
\begin{align}
m_{3/2} &= e^{K/2}\abs{W} \sim \tau^2
\label{eqn:gravitino}
\end{align}
implying that $\tau^2$ sets the scale of the gravitino mass (and also the mass of the Polonyi field). However $\tau^2\sim\Upsilon e^{-2aS}$, consequently the gravitino mass originates from the gaugino condensation and is doubly suppressed. A concrete realization based on the hidden sector gauge group $SU(8)$ is shown in fig.\,\ref{fig:benchmark} and tab.\,\ref{tab:sample}, summarizing the main parameters.

\begin{table}
\small
\begin{center}
\begin{tabular}{ v{1.5cm}v{1.5cm}v{0.8cm}v{1.5cm}v{1.5cm}v{1.1cm}v{1.5cm}v{1.1cm} }
\midrule\addlinespace
$\nu$ & $\tau^2$ & $C_0$ & $F^S$ & $F^C$ & $m_{3/2}$ & $m_S$ & $m_C$ \\\cmidrule(lr){1-8}
$4\times10^{-15}$ & $2\times10^{-14}$ & $0.73$ & $9\times10^{-16}$ & $2\times10^{-14}$ & $\unit[25]{TeV}$ & $\unit[1909]{TeV}$ & $\unit[43]{TeV}$\\\addlinespace
\midrule\midrule
\end{tabular}
\end{center}
\caption{Sample spectrum with a multi-TeV gravitino for $SU(8)$.}
\label{tab:sample}
\end{table}

To analyze the structure of the resulting soft terms we focus on the case $C_0\ll 1$. Allowing couplings between hidden matter $C$ and observable fields $Q_\alpha$ in the K\"ahler potential
\begin{align}
K &= -\log{\klammer{S+\ol{S}}} + C\ol{C} + Q_\alpha\ol{Q}_\alpha \Big[ 1 + \varsigma_\alpha C\ol{C} \Big],
\label{eqn:fullkahler}
\end{align}
the emerging soft terms are given by
\begin{align}
m^2_\alpha &= \varsigma_\alpha \frac{\abs{F^S}^2}{\klammer{S_0+\ol{S}_0}^2} + \klammer{1-3\varsigma_\alpha}m^2_{3/2} + \mathtt{ANOMALY} , \label{eqn:softscalars} \\
A_{\alpha\beta\gamma} &= -\frac{F^S}{S_0 + \ol{S}_0} + \mathtt{ANOMALY}, \label{eqn:softaterms} \\
M_a &= \frac{F^S}{S_0 + \ol{S}_0} + \mathtt{ANOMALY}, \label{eqn:softgauginos}
\end{align}
where ``\texttt{ANOMALY}'' denotes possible 1-loop and 2-loop
contributions to the soft terms. As eq.\,(\ref{eqn:fs}) shows, the dilaton contribution is suppressed by the little hierarchy.
Therefore gaugino
masses and $A$-terms feel the contribution from dilaton and anomaly
in comparable portions in the spirit of a hybrid mediation, also
know as \emph{mirage mediation} \cite{Choi:2004sx,Choi:2005ge,Choi:2005uz,Falkowski:2005ck,Baer:2006id,Endo:2005uy,Lebedev:2006tr,Lebedev:2006qq,GomezReino:2006dk,Lust:2006zg,Dudas:2006gr,Abe:2006xp,Lebedev:2006qc,LoaizaBrito:2005fa}. The fate of the soft scalar
masses is more model dependent and
crucially depends on the parameter $\varsigma_\alpha$ which
describes the coupling between hidden and observable
matter.
For $\varsigma_\alpha\ll1$ the dilaton as well as the anomaly
contribution are negligible. In this scenario the scalar masses
are dominated by the $F$-term of the downlifting field $C$.
On the other hand, for $\varsigma_\alpha\sim\mathcal{O}(\nicefrac{1}{3})$
the matter contribution to scalar masses becomes suppressed
whereas dilaton and anomaly mediated parts provide comparable contributions.

While one can realize different scenarii for the scalar masses,
mirage mediation for the gauginos and the $A$-terms
is more reliable and independent
of $\varsigma_\alpha$. The gaugino masses encode the mirage
unification feature in the most robust way \cite{Choi:2007ka}. This mirage scale is determined by the
relative strength of dilaton versus anomaly mediation in
eq.\,(\ref{eqn:softgauginos}). Fig.\,\ref{fig:gauginos} shows
the evolution of the gaugino masses for equal dilaton/anomaly
contribution. Finally, the masses of the dilaton, the Polonyi
field, the gravitino and the gauginos exhibit the hierarchy
\begin{align}
m_S \sim \Upsilon m_{3/2} \sim \Upsilon m_C \sim \Upsilon^2 m_{1/2}.
\end{align}

\section{Applications}

Let us try to see how the new scenario with a doubly suppressed
gravitino mass can be applied to various aspects of model building
in particle physics and cosmology.

\subsection[Small gravitino mass with a large $\Lambda$]{Small gravitino mass with a large $\boldsymbol{\Lambda}$}

The standard formula \cite{Nilles:1982ik} for the gravitino mass from a
gaugino condensate reads
\begin{align}
m_{3/2} \sim \frac{\Lambda^3}{M_\s{P}^2}
\end{align}
and $F\sim \Lambda^3 / M_\s{P}$. It requires $\Lambda$ to be at an
intermediate scale if $m_{3/2}$ is at the (multi) TeV scale.
To raise $\Lambda$ in a
realistic set-up would require soft terms that are small
compared to $m_{3/2}$ \cite{Nilles:1982ik} and this is not easy to achieve.
The doubly suppressed solution gives
\begin{align}
m_{3/2} \sim \frac{\Lambda^6}{M_\s{P}^5},
\end{align}
and avoids an intermediate scale.

In a rather natural way $\Lambda$ could be identified with
the grand unified scale $M_\s{GUT}$ or the compactification
scale $M_\s{COMP}$ of extra
dimensions in string theory, typically assumed to be at
a scale of few times $\unit[10^{16}]{GeV}$. Thus a single scale $\Lambda$ might
represent $M_\s{GUT} \sim  M_\s{COMP}$ as well as the hierarchically
small scale $m_{3/2}$. Model building along these limes might be promising.

In our benchmark model we obtained a hidden sector group $SU(8)$ assuming
a pure supersymmetric $SU(N)$ gauge theory as well as the equality of the
gauge coupling constants of hidden and observable sector. This group could
originate certainly from the $SO(32)$ theory but not so easily \cite{Lebedev:2006tr} from the $E_8\times E_8$ theory
favored by phenomenological arguments \cite{Lebedev:2006kn,Lebedev:2007hv}.
String threshold corrections, however, might
enlarge the hidden sector gauge coupling compared to the observable sector one, allowing for smaller hidden sector gauge groups and thus
reopening many new ways for model building.
In fact, in heterotic M-theory \cite{Horava:1995qa,Horava:1996ma},
a larger coupling in the hidden sector might appear in a natural way \cite{Witten:1996mz,Nilles:1997cm,Nilles:1998sx}.
Such models might then explain all scales directly from the string scale,
without invoking the existence of an intermediate scale.

\subsection[The $\mu$-problem in gauge mediation]{The $\boldsymbol{\mu}$-problem in gauge mediation}

If we considered a situation with condensation at the intermediate
scale in a model with double suppression we would obtain a rather smallish
gravitino mass. SUSY breaking at the weak scale would then need
another source,
as in models of supersymmetry breakdown via gauge mediation \cite{Dine:1981za,Dimopoulos:1981au,Giudice:1998bp} (i.\,e. the gravitino
mass is small compared to the soft SUSY breaking terms).
Here one of the challenges is
the generation of the $\mu$-term $\mu H_1H_2$ for the Higgsino masses at a
scale
comparable to the soft terms. In such a model we could now consider a
condensate at an intermediate scale and a term in the superpotential \cite{Chun:1991xm}
\begin{align}
\frac{1}{M_\s{P}} Q\overline{Q}H_1H_2,
\end{align}
where $Q, \overline{Q}$ denotes hidden quark superfields. A condensation at
the intermediate scale would then lead to an effective $\mu$-term in the
TeV-region. In a model with a doubly suppressed gravitino mass,
$m_{3/2}$ would then be at a scale of the order of $\unit[10^{-3}]{eV}$.

\subsection{Composite axions for dark energy and inflation}

Axions are promising candidates for quintessence fields as
they only have derivative couplings and thus avoid some of the problematic
aspects of light (real) scalar fields. They frequently appear in string theory and often have decay constants of order of the Planck
scale. A specific scheme for a (composite) quintaxion has been outlined
in \cite{Kim:2002tq}. As the
vacuum energy is small $(\unit[10^{-3}]{eV})^4$ the axion potential has to be
extremely flat and the quintaxion mass exceedingly small.
As pointed out in \cite{Kim:2002tq} such a situation can be realized in the presence of
a strongly interacting hidden sector with (almost) massless hidden
quarks. With massless quarks the $\theta$-angle is unphysical and the axion
potential is flat. Parametrically the vacuum energy is given by
\begin{align}
\lambda^4 \sim m_Q^n m_g^N \Lambda^{(4-n-N)},
\end{align}
where $m_Q$ is the mass of (n quarks) and $m_g$ is the gluino
mass of an $SU(N)$ gauge group. In the presence of massless quarks and or
unbroken supersymmetry
$\lambda=E_\s{VAC}=0$. The key motivation for the model in \cite{Kim:2002tq} was the fact
that hidden sector quarks played a crucial role in the generation of
$\mu$-term in the superpotential ($\mu H_1H_2$) of the
supersymmetric standard model, through higher order
couplings in the superpotential, like the one discussed previously
\begin{align}
\frac{1}{M_\s{P}} Q\overline{Q}H_1H_2.
\end{align}
Once the Higgs fields receive nontrivial vacuum expectation
values, these same terms induce a hierarchically tiny mass
for the hidden sector quarks in a kind
of gravitational see-saw mechanism involving the weak scale and the Planck
scale. This then made it possible to obtain an axion potential sufficiently
flat to be compatible with a realistic value of $\lambda=E_\s{VAC}$,
although many details of model building have to be clarified \cite{Kim:2006aq}.
One might also include a second invisible axion that solves the
$\theta$-problem of
QCD. The model in \cite{Kim:2002tq} considered an intermediate scale
$\Lambda \sim \unit[10^{13}]{GeV}$ (responsible for
$m_{3/2} \sim \Lambda^3 /M^2_\s{P}$).
Models with a doubly suppressed gravitino mass and a scale
$\Lambda \sim M_\s{GUT}$ provide a novel way to reconsider
quintessential axions, although in a somewhat modified set-up.
Here we need a contribution to the $\mu$ term that is more strongly
suppressed than the one considered above. Still the fact remains true,
that such
a term (relevant for $\mu$) would also induce a tiny hidden quark
mass term and thus a tiny contribution to the vacuum energy
once the Higgs fields receive a nonvanishing VEV.
A detailed discussion is beyond the scope of this paper and will be
addressed in future work. Needless to say, that these consideration
are far from a solution to the cosmological constant problem, as
other contributions to the vacuum energy (as e.\,g. from SUSY breakdown
or from electroweak symmetry breakdown) have to vanish.

Axions might also find important implications in the study of
models for the inflationary universe \cite{Freese:1990rb,Adams:1992bn,Kim:2004rp,Dimopoulos:2005ac,Grimm:2007hs}. The requirement there
is that the effective axion decay constant should be rather large. Again the
models considered here with a composite axion in a doubly
suppressed solution might open new aspects for model building.

\subsection*{Acknowledgements}

We thank Jihn E. Kim, Costas Kounnas, Oleg Lebedev, Andrei Micu, Marios
Petropoulos, Marco Serone and Patrick Vaudrevange for useful
conversations.

This work was partially supported by the
European Union 6th framework program MRTN-CT-2004-503069
``Quest for unification'', MRTN-CT-2004-005104 ``ForcesUniverse'',
MRTN-CT-2006-035863 ``UniverseNet'' and
SFB-Transregio 33 ``The Dark Universe'' by Deutsche
Forschungsgemeinschaft (DFG).

\addcontentsline{toc}{section}{{Bibliography}}
\providecommand{\href}[2]{#2}\begingroup\raggedright
\endgroup

\end{document}